%% file: main.tex
\documentclass[conference]{IEEEtran}
\IEEEoverridecommandlockouts

\usepackage{multicol}
\usepackage{amsmath,amssymb,amsfonts}
\usepackage{algorithmic}
\usepackage[pdftex]{graphicx}
\usepackage{textcomp}
\usepackage{xcolor}
\usepackage{tcolorbox}

\usepackage{subfigure}
\usepackage{multirow}
\usepackage{booktabs}
\usepackage{marvosym}
\usepackage{xcolor} 

\usepackage{soul}

\usepackage{enumitem}
\setitemize{noitemsep,topsep=0pt,parsep=0pt,partopsep=0pt,leftmargin=2mm}
\usepackage{url}

\usepackage[normalem]{ulem}
\useunder{\uline}{\ul}{}
\newcommand{\stitle}[1]{\smallskip\noindent{\bf #1}}

\let\oldbibliography\thebibliography
\renewcommand{\thebibliography}[1]{\oldbibliography{#1}
\setlength{\itemsep}{0pt}} 

\usepackage[labelsep=period]{caption}
\renewcommand{\thetable}{\arabic{table}}

\newtheorem{exmp}{Example}[section]

\def\BibTeX{{\rm B\kern-.05em{\sc i\kern-.025em b}\kern-.08em
    T\kern-.1667em\lower.7ex\hbox{E}\kern-.125emX}}
\begin{document}


\title{CleanML: A Study for Evaluating the Impact of Data Cleaning on ML Classification Tasks
}


\author{
Peng Li$^{\dagger}$, Xi Rao$^{\ddagger}$,
Jennifer Blase$^{\dagger}$, Yue Zhang$^{\dagger}$, Xu Chu$^{\dagger}$,
Ce Zhang$^{\ddagger}$ \\
$^\dagger$\textit{Georgia Institute of Technology}, $^\ddagger$\textit{ETH Zurich} \\
$^\dagger$\{pengli, jblase, yzhang3271, xu.chu\}@gatech.edu, $^\ddagger$\{rao, ce.zhang\}@inf.ethz.ch}

\setcounter{page}{1}
\maketitle

\begin{abstract}

Data quality affects machine learning (ML) model performances, and data scientists spend considerable amount of time on data cleaning before model training. However, to date, there does not exist a rigorous study on how exactly cleaning affects ML --- ML community usually focuses on developing ML algorithms that are robust to some particular noise types of certain distributions, while database (DB) community has been mostly studying the problem of data cleaning alone without considering how data is consumed by downstream ML analytics. 

We propose a CleanML study that systematically investigates the impact of data cleaning on ML classification tasks. The open-source and extensible CleanML study currently includes 14 real-world datasets with real errors, five common error types, seven different ML models, and multiple cleaning algorithms for each error type (including both commonly used algorithms in practice as well as state-of-the-art solutions in academic literature). We control the randomness in ML experiments using statistical hypothesis testing, and we also control false discovery rate in our experiments using the Benjamini-Yekutieli (BY) procedure. We analyze the results in a systematic way to derive many interesting and nontrivial observations. We also put forward multiple research directions for researchers. 
\end{abstract}

\input{introduction}
\input{related_work}
\input{cleanml_db_schema}

\input{cleanml_db_instance}
\input{cleanml_db_analysis}
\input{conclusion}

\vspace{-1mm}
\section{Acknowledgement}
\vspace{-1mm}
The Chu Data Lab acknowledges the support from SCS in GT, Georgia State funds, as well as the JP Morgan Faculty award.
CZ and the DS3Lab gratefully acknowledge the support from the Swiss National Science Foundation (Project Number 200021\_184628), Innosuisse/SNF BRIDGE Discovery (Project Number 40B2-0\_187132), European Union Horizon 2020 Research and Innovation Programme (DAPHNE, 957407), Botnar Research Centre for Child Health, Swiss Data Science Center, Alibaba, Cisco, eBay, Google Focused Research Awards, Oracle Labs, Swisscom, Zurich Insurance, Chinese Scholarship Council, and the Department of Computer Science at ETH Zurich.

\bibliographystyle{abbrv}
\bibliography{vldb_sample}
\end{document}

%% file: introduction.tex

\section{Introduction}
\label{sec:introduction}

The quality of machine learning (ML) applications is only as good as the quality of the data it trained on, and data cleaning has been the cornerstone of building high-quality ML models. Not surprisingly, both ML and database (DB) communities have been working
on problems associated with dirty data:
\begin{itemize}[noitemsep]
\item {\bf ML community} has been focusing on understanding the impact of noises on ML models without actually performing data cleaning. 
On the one hand, many ML models are robust to small amounts of random noises --- there has been research showing that the noise introduced during the training process (e.g., via asynchronous communication and lossy compression) can have negligible effect in accuracy, both empirically and theoretically~\cite{alistarh2017qsgd, de2017understanding, lian2017can, recht2011hogwild, zhang2017zipml, zhang2015staleness}. 
On the other hand, ML models can also be sensitive to other types of noises, especially those non-white noises that are in the input data~\cite{krishnan2016activeclean} and labels~\cite{frenay2014classification}. Instead of performing data cleaning, the ML community has mostly been focusing on designing ML algorithms that are {\em robust} to noises of certain distributions, such as noise-robust decision trees~\cite{quinlan1987simplifying}, the use of regularization for improving robustness~\cite{teng2000evaluating}, and model bagging to reduce the variability of model performances caused by dirty data~\cite{khoshgoftaar2011comparing}. 

\item {\bf DB community} has been mostly focusing on understanding the fundamental process of data cleaning without considering its impact on ML models. 
Data cleaning usually consist of two phases: \textit{error detection}, where various errors are identified and possibly validated by experts; and \textit{error repair}, where updates to the database are applied (or suggested to human experts) to make the data cleaner. Many techniques have been proposed for detection, for example, by designing integrity constraints to capture data inconsistencies~\cite{chu2013discovering}, by using statistical techniques to detect outliers~\cite{hellerstein2008quantitative}, and by building ML models to detect duplicates~\cite{elmagarmid2007duplicate}. Various techniques have also been proposed for repairing, for example, by finding the minimal set of updates to resolve violations~\cite{chu2013holistic}, by performing data transformations~\cite{he2018transform}, by consulting external knowledge bases~\cite{chu2015katara}, and by using probabilistic graphical models to reason about errors holistically~\cite{rekatsinas2017holoclean}.
\end{itemize}



\begin{figure}[t!]
    \centering
    \includegraphics[width=0.95\columnwidth]{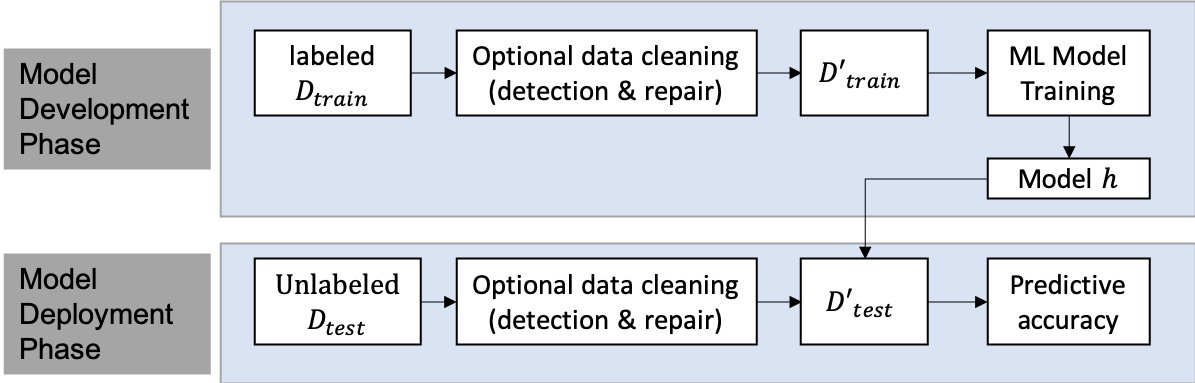}
    \vspace{-2mm}
    \caption{\scriptsize Typical ML workflow with data cleaning.}
    \label{fig:cleanml_workflow}
    \vspace{-6mm}
\end{figure}

\noindent
{\bf (Goal of This Work.)}
In real applications, these two angles are less
segregated. Indeed, data cleaning is often seen as a crucial data preparation step either before training an ML model on a labeled training set in the model development phase, or before making predictions on an unlabeled test set in the model deployment phase~\cite{polyzotis2017data} (Figure~\ref{fig:cleanml_workflow}). It is reported that data  scientists reportedly spend up to 80\% of their time performing various data cleaning activities~\cite{cleaningdatatimeconsuming}.

However, limited prior work exists in studying the impact of data cleaning for downstream ML model performance, and they tend to focus on some specific error types (e.g., domain value errors in BoostClean~\cite{krishnan2017boostclean}), specific cleaning methods (e.g., human oracle for cleaning in ActiveClean~\cite{krishnan2016activeclean}), and/or specific ML models (e.g., convex models trained using stochastic gradient descent in ActiveClean~\cite{krishnan2016activeclean} and a weighted ensemble model in BoostClean~\cite{krishnan2017boostclean}). 


The goal of this paper is not to propose a new approach of data cleaning for ML. Instead, our goal is to (1) conduct a first systematic empirical study on the impact of data cleaning on downstream ML classification models, for different error types, cleaning methods, and ML models; 
(2) given our empirical findings, provide a starting point for future research to advance the field of cleaning for ML.

\noindent
{\bf (Challenges)} Conducting such a systematic study is not a trivial task with the following challenges:

\begin{itemize}[noitemsep]

\item {\bf Comprehensive Study Scope.} As shown in Figure~\ref{fig:cleanml_workflow}, the accuracy of an ML model depends on the dataset, the ML model, whether data cleaning is applied, where data cleaning is applied (on training data or on test data), and what cleaning algorithm is used. Studying a particular aspect of the workflow (e.g., logistic regression as~\cite{sarkar2011detection}) is not enough to provide a qualitative assessment on the general impact of data cleaning. We thus need a principled way to organize different experiments and analyze results systematically.

\item {\bf Realistic Error Patterns.} Most of the current data cleaning work in the DB community uses synthetically injected errors (and hence the ground-truth clean data is known) to evaluate data cleaning algorithms~\cite{arocena2015messing}. 
Taking a standard ML dataset with simulated data fallacies (e.g.,  by randomly removing values to mimic missing values) might under/over-estimate the impact of data cleaning on ML. 
For our study to reflect the real-world impact of data cleaning on ML, we have to work with real-world datasets that contain realistic errors, for which we usually do not have ground-truth clean versions.

\item {\bf Statistically Significant Results and Controlling False Discoveries.} ML models are inherently probabilistic, for example, a different train/test split on the same dataset
might produce entirely different results. Therefore, ensuring the statistically significance of our findings presents a major challenge. This is further complicated by the fact that we have many different datasets, ML models, and data cleaning choices to cover, which will likely produce false discoveries,  known as  the multiple hypothesis testing problem~\cite{rupert2012simultaneous,zhao2017controlling}.   


\end{itemize}
\noindent
{\bf (Main Contributions of CleanML.)} We make the following contributions in our proposed CleanML study.

\begin{itemize}[noitemsep]
\item {\bf CleanML Scope of Study.} We focus on \emph{five types of errors}, including outliers, duplicates, inconsistencies, mislabels and missing values. These errors are prevalent in the real-world datasets and frequently considered in research. We select \emph{seven classification algorithms}, including Logistic Regression, Decision Tree, Random Forest, Adaboost, XGBoost, k-Nearest Neighbors (KNN) and Naive Bayes. These are classical and competitive classification models commonly used on classification tasks on structured datasets. We collected \emph{14 real-world datasets} containing different types of errors, and devised meaningful classification tasks on the datasets. We organize all experimental results in a relational database named CleanML, whose schema we describe in Section~\ref{sec:cleanml_schema}. 

\item {\bf Obtaining Cleaned Versions of Dirty Datasets.} As we do not have ground-truth versions of the dirty datasets, for each error type, we selected various automatic data cleaning algorithms, including both simple data cleaning solutions commonly used in practice (e.g., mean imputation), as well as state-of-the-art data cleaning solutions proposed in academic literature (e.g., probabilistic cleaning solution HoloClean~\cite{rekatsinas2017holoclean} and unsupervised duplicate detection solution ZeroER~\cite{wu2020zeroer}). We also include a study comparing human cleaning with automatic cleaning.


\item {\bf Controlling Randomness and False Discoveries.} We describe the steps taken to populate the CleanML database instance in Section~\ref{sec:cleanml_instance}, where we controlled randomness by using multiple train/test splits to conduct a statistical hypothesis testing procedure, the paired sample $t$-test~\cite{mcdonald2009handbook}. We also leveraged the Benjamini-Yekutieli (BY) procedure \cite{friedman2001elements} to control false discovery rate.

\item {\bf Empirical Findings and Future Research Directions.} We obtain many interesting and non-trivial empirical findings by systematically issuing various queries against the CleanML relational database. We present a detailed analysis for each error type in Section~\ref{sec:cleanml_analysis}. We summarize our findings in Section~\ref{sec:conclusion}. We also put forward multiple directions for future research in the area of cleaning for ML in Section~\ref{sec:future_research}.


\item {\bf Extensible Open-source Code.}  We made the code and datasets  publicly available for reproducibility~\footnote{\scriptsize \url{https://chu-data-lab.github.io/CleanML/}}. In addition, our CleanML study can be easily extended by adding new datasets, error types, cleaning algorithms, or ML models --- the code for running  experiments and for performing result analysis can be reused without  modification.
\end{itemize}

%% file: related_work.tex
\vspace{-3mm}
\section{Related Work}
\vspace{-3mm}
\stitle{Data Cleaning.} Despite many years of research, the dirty data problem remains challenging~\cite{chu2016data}. Recent study shows that even the combination of current error detection techniques can still miss many errors in real-world datasets~\cite{abedjan2016detecting}. We refer readers to various surveys and tutorials on the broad topic of data cleaning~\cite{rahm2000data,hellerstein2008quantitative,fan2012foundations,ilyas2015trends}. The recent efforts on data cleaning focus on leveraging advanced ML techniques, including the HoloDetect system that uses few-shot learning and data augmentation for error detection~\cite{DBLP:conf/sigmod/HeidariMIR19}, the HoloClean system that uses probabilistic graph models for inferring the most likely values~\cite{rekatsinas2017holoclean}, deep-learning based methods for duplicate detection~\cite{mudgal2018deep,ebraheem2018distributed}, and the ZeroER that uses generative model for detecting duplicates with zero labeled examples~\cite{wu2020zeroer}. 

\stitle{Analytics-Driven Cleaning.} As data cleaning itself is expensive and hard to reach ground truth, the DB community is starting to work on analytics-driven cleaning methods. SampleClean~\cite{wang2014sample} targets the problem of answering SQL aggregate queries when the input data is dirty by cleaning a sample of the dirty dataset, and at the same time, providing statistical guarantees on the query results.
ActiveClean~\cite{krishnan2016activeclean} is an example of cleaning data  for convex ML models that are trained using gradient descent. The key insight of ActiveClean is that convex loss models (e.g., linear regression) can be trained and cleaned simultaneously using mini-batch gradient descent. ActiveClean assumes that the cleaning is performed by a human oracle while CleanML evaluates automatic cleaning algorithms. BoostClean~\cite{krishnan2017boostclean} automatically selects from a predefined space of cleaning methods using a hold-out validation set via statistical boosting. BoostClean only considered \textit{domain value violations} when an attribute value is outside of its value domain and only tested random forest model, while CleanML considers five error types and seven ML models.

%% file: cleanml_db_schema.tex
\vspace{-2mm}
\section{CleanML Database Schema}
\label{sec:cleanml_schema}


As shown in Figure~\ref{fig:cleanml_workflow}, there are multiple factors contributing to the impact of data cleaning on ML classification tasks. To effectively organize the study, we propose to use a relational database (named the CleanML database) to store the results of different experiments. We first discuss the three relations in the CleanML database in Section~\ref{sec:cleanml_three_relations}. We then present the domain of each attribute in the relations from Sections~\ref{sec:cleanml_cleaning_attributes} to~\ref{sec:cleanml_scenarios}, which together define the scope of our study. 

\subsection{The Three Relations}
\vspace{-4mm}
\label{sec:cleanml_three_relations}
\begin{table}[!h]
\caption{\scriptsize{CleanML  Schema. Keys are underlined.}}
\vspace{-2mm}
\scalebox{0.8}{
\begin{tabular}{lllllll}
\multicolumn{7}{l}{\textbf{R1 (Vanilla)}} \\ \hline
\multicolumn{1}{|l|}{\ul{Dataset}} & \multicolumn{1}{l|}{\ul{Error Type}} & \multicolumn{1}{l|}{\ul{Detection}} & \multicolumn{1}{l|}{\ul{Repair}} & \multicolumn{1}{l|}{\ul{ML Model}} & \multicolumn{1}{l|}{\ul{Scenario}} & \multicolumn{1}{l|}{Flag} \\ \hline
\end{tabular}
}
\\\\
\scalebox{0.8}{
\begin{tabular}{llllll}
\multicolumn{6}{l}{\textbf{R2 (With Model Selection)}} \\ \hline
\multicolumn{1}{|l|}{\ul{Dataset}} & \multicolumn{1}{l|}{\ul{Error Type}} & \multicolumn{1}{l|}{\ul{Detection}} & \multicolumn{1}{l|}{\ul{Repair}} & \multicolumn{1}{l|}{\ul{Scenario}} & \multicolumn{1}{l|}{Flag} \\ \hline
\end{tabular}
}
\\\\
\scalebox{0.8}{
\begin{tabular}{lllll}
\multicolumn{5}{l}{\textbf{R3 (With Model Selection and Cleaning Method Selection)}} \\ \cline{1-4}
\multicolumn{1}{|l|}{\ul{Dataset}} & \multicolumn{1}{l|}{\ul{Error Type}} & \multicolumn{1}{l|}{\ul{Scenario}} & \multicolumn{1}{l|}{Flag} &  \\ \cline{1-4}
\end{tabular}
}
\label{dcml_schema}
\end{table}
\vspace{-2mm}





The CleanML relational schema consists of three relations, as shown in Table~\ref{dcml_schema}. The primary keys of all three relations are underlined, respectively. We first introduce the attributes, and then highlight the differences between these three relations. 

\begin{itemize}[noitemsep]
\item \textbf{Dataset Attribute.} The first attribute is \textit{dataset}, which specifies a dataset under study. Each dataset has an associated ML task. 
Each dataset can have multiple types of errors.
Instead of injecting synthetic errors into datasets, we use real-world datasets with real errors, and we apply various cleaning methods to detect and repair the errors. We list all the datasets we use in Section~\ref{sec:cleanml_datasets}.



\item \textbf{Attributes for Data Cleaning.} 
The \textit{error type} attribute specifies the error type under consideration. We include five most common types of dirtiness considered in the ML and DB communities: missing values, outliers, duplicates, inconsistencies and mislabels. 
For each error type, we consider the most commonly used cleaning methods in practice as well as the state-of-the-art cleaning algorithm in academic literature. Each cleaning method includes an error \textit{detection} component and an error \textit{repair} component. We discuss the cleaning methods considered  in Section~\ref{sec:cleanml_cleaning_attributes}. We not only look at single error cleaning, but  also discuss mixed error cleaning in Section \ref{sec:mixed_errors}.


\item \textbf{Attribute for ML Model.} 
The \textit{ML model} specifies the ML algorithm used for classification. Different ML models may have different behaviors to different error types. We describe the chosen ML models in Section~\ref{sec:cleanml_models}.

\item \textbf{Attribute for Cleaning Scenario.} The \textit{scenario} attribute indicates whether cleaning is applied in the training set or the test set (c.f. Figure~\ref{fig:cleanml_workflow}). In other words, it specifies whether we are evaluating the impact of data cleaning on ML in the model development phase or in the model deployment phase. We explain this further in Section~\ref{sec:cleanml_scenarios}.

\item  \textbf{Flag Attribute.} The \textit{flag} attribute summarizes the impact of data cleaning on ML for a specific setting under consideration, i.e., a particular combination for the key attributes. It has three possible values: ``P (positive)'', ``N (negative)'' or ``S (insignificant)'', indicating the cleaning has positive, negative, or insignificant impact on the ML performance, respectively. 
\end{itemize}

\stitle{Relation $R1$.} $R1$ is the vanilla version of our CleanML relations. Its primary key is \textit{dataset, error type, detection, repair, ML model, scenario}. Every tuple of $R1$ stores the result (flag) of a specific experiment: \textit{how does cleaning some type of error using a detection method and a repair method affect a ML model for a given dataset?} 

\stitle{Relation $R2$.} Given an ML classification task, ML developers often  perform model selection to select the best model (e.g., by using cross-validation). Compared with $R1$, $R2$ eliminates the ML model attribute. In this case, we try different models during training and select the model that has the best validation accuracy (or F1 score if the dataset has imbalanced classes) as the model to be considered in an experiment in $R2$. Every tuple of $R2$ stores the result (flag) of a specific experiment with model selection: \textit{how does cleaning some type of error using a detection method and a repair method affect the best ML model for a given dataset?} 
Note that we must create a new relation $R2$ for this question, instead of simply picking one tuple out of the tuples in $R1$ with the same dataset, error type, detection, repair, and scenario attribute values. This is because every tuple in $R1$ is actually an aggregation over different train/test splits in order to control randomness in the experiments (c.f., Section~\ref{sec:control_randomness}), and the best model for each train/test split might be different. 

\stitle{Relation $R3$}. Compared with $R2$, $R3$ further eliminates the cleaning method (detection and repair) attributes. In this case, in addition to model selection, we also try different cleaning methods and select the one that results in the best validation accuracy (or F1 score if the dataset has imbalanced classes). This is essentially the cleaning algorithm selection strategy employed in BoostClean~\cite{krishnan2017boostclean}.
Every tuple of $R3$ stores the result (flag) of a specific experiment with model selection and cleaning method selection: \textit{how does the best cleaning method affect the performance of the best model for a given dataset?} 
Again, we must create a new relation $R3$ for addressing this question, instead of picking one tuple from $R2$ with the same dataset, error type, and scenario attribute values. This is because every tuple in $R2$ is also an aggregation over different train/test splits (c.f., Section~\ref{sec:control_randomness}), and the best cleaning method for each split may be different.




\subsection{Error Types and Automatic Cleaning Methods}
\label{sec:cleanml_cleaning_attributes}

We consider five error types prevalent in the real-world datasets, including missing values, outliers, duplicates, inconsistencies and mislabels. For each error type, we consider  various automatic  cleaning methods commonly used in practice as well as the state-of-the-art cleaning methods in academic literature. Note that all algorithms discussed in this section are automatic cleaning algorithms that require no or minimal human interactions. We will separately study human cleaning in Section~\ref{sec:human_vs_automatic}. All automatic cleaning methods are described below and listed in Table \ref{clean_methods}.

\vspace{-1mm}
\begin{table}[!h]
\centering
\caption{\scriptsize Automatic Cleaning Methods}
\resizebox{0.9\linewidth}{!}{
\vspace{-3mm}
\begin{tabular}{c|c|c}
\toprule
\multicolumn{1}{c|}{\textbf{Error Type}} & \multicolumn{1}{c|}{\textbf{Detection Method}} & \multicolumn{1}{c}{\textbf{Repair Method}} \\ 
\midrule
\multirow{5}{*}{Missing Values} & \multirow{5}{*}{\begin{tabular}[c]{@{}l@{}}Empty \\ Entries\end{tabular}} & Deletion \\ \cline{3-3} 
 &  & \hspace{6pt} Mean\_Mode, Mean\_Dummy \\
 &  & \hspace{6pt} Median\_Mode, Median\_Dummy \\
 &  & \hspace{6pt} Mode\_Mode, Mode\_Dummy \\ \cline{3-3} 
 &  & HoloClean \\ \hline
\multirow{3}{*}{Outliers} & SD &\multirow{2}{*}{ Mean, Median, Mode}  \\ \cline{2-2}
 & IQR & \hspace{6pt} \\ \cline{2-3} 
 & IF & HoloClean \\ \hline
\multirow{2}{*}{Duplicates} & Key Collision & \multirow{2}{*}{Deletion} \\ \cline{2-2}
 & ZeroER &  \\ \hline
Inconsistencies & OpenRefine & Merge \\ \hline
Mislabels & cleanlab & cleanlab \\ 
\bottomrule
\end{tabular}
\vspace{-5mm}}
\label{clean_methods}
\vspace{-4mm}
\end{table}

\vspace{0.3em}
\subsubsection{Missing Values}
 Missing values occur when no value is stored for cells. Missing values can be naturally detected by finding empty or $NaN$ (a commonly used placeholder) entries.
 We use the following methods to repair missing values:
\begin{itemize}[noitemsep]
\vspace{-1mm}
\item \textbf{Deletion:} Delete records with missing values.
\item \textbf{Six Ways of Simple Imputation:} For numerical missing values, we consider three types of imputation methods: mean, median and mode. For categorical missing values, we use two types of imputation methods: the mode (most frequent class) or a dummy variable named ``missing". Therefore, we have six imputation methods. In Table \ref{clean_methods}, we denote each imputation method by the numerical imputation and categorical imputation (e.g. ``Mean\_Dummy" represents imputing numerical missing values by mean and imputing categorical missing values by dummy variables).
\item \textbf{Probabilistic Inference (HoloClean):} HoloClean~\cite{rekatsinas2017holoclean} is a state-of-the-art statistical inference engine to impute, clean, and enrich data. It leverages multiple signals (e.g., value correlations and any available reference data) holistically to build a probabilistic model for inferring what is the most likely value for a given cell.

\end{itemize}

\subsubsection{Outliers} An outlier is an observation that is distant from others~\cite{DBLP:books/sp/Aggarwal2013}. We detect numerical outliers as follows:
\begin{itemize}[noitemsep]
    \item \textbf{Standard Deviation Method (SD):} A value in a column is considered to be an outlier if it is $n$ numbers of standard deviations away from the mean. We use $n=3$.
    \item \textbf{Interquartile Range Method (IQR):} Let $Q_1$ and $Q_3$ be the 25th and the 75th percentiles of an attribute. Then, the interquartile range $IQR = Q_3 - Q_1$. A value is considered to be an outlier if it is outside the range of [$Q_1-k\times IQR$, $Q_3+k\times IQR$]. We use $k=1.5$.
    \item \textbf{Isolation Forest Method (IF):} The isolation forest isolates observations by randomly selecting a feature and a split value of the selected feature. This partition can be represented by a tree structure and it produces noticeably shorter paths for outliers. We use the scikit-learn IsolationForest~\cite{scikit-learn} and set the contamination parameter to be $0.01$.
\end{itemize}
We repair outlying cells using simple imputation and HoloClean, which are exactly the same as those for repairing missing values, except that we only have three ways of imputation since we consider only numerical outliers.



\subsubsection{Duplicates} Duplicates refer to the records that correspond to the identical real-world entity~\cite{elmagarmid2007duplicate}. We consider a simple method commonly used in practice as well as a state-of-the-art unsupervised approach for detecting duplicated records:
\begin{itemize}[noitemsep]
    \item \textbf{Key Collision:} This method uses the key attributes that are supposed to be unique for every tuple in a dataset, and declares two records as duplicates if they have the same value on the key attributes.
    \item \textbf{Unsupervised ER (ZeroER):} While there are many supervised approaches for detecting duplicates, including recent deep learning based methods~\cite{mudgal2018deep,ebraheem2018distributed}, real-world datasets rarely come with labeled duplicates and non-duplicates. Hence, we choose a state-of-the-art unsupervised method, called ZeroER~\cite{wu2020zeroer}, that achieves comparable performance to supervised methods but requires zero labeled examples.
\end{itemize}
For a set of records that are deemed to be duplicates, we repair them by deleting all but one record in the set. 

\subsubsection{Inconsistencies} Inconsistencies occur when two cells in a column have different values, but should actually have the same value~\cite{ilyas2015trends}. For example, both ``CA'' and ``California'' can appear together in a state column. We use a popular open-source tool for detecting inconsistencies: 

\begin{itemize}[noitemsep]
    \item \textbf{OpenRefine:}  \textit{OpenRefine} \cite{verborgh2013using}  provides a \textit{text facets clustering} function, which is able to find groups of different values that might be alternative representations of the same thing. For example, after clustering text facets on the company name attribute, ``U.S. Bank" and ``US Bank" will be clustered into the same group and can be identified as inconsistencies. We use this function to detect inconsistencies. Fixing inconsistencies can be done by simply merging all values in one cluster into the most frequent one.
\end{itemize}




\subsubsection{Mislabels} \label{subsec:mislabels}
 Mislabels occur when an example is incorrectly labeled. As we only have one dataset (Clothing) with real mislabels, we perform synthetic mislabel injection on additional four datasets (c.f. Table~\ref{T:dataset})
, following the strategies in \cite{garcia2015data}: (1) \textit{uniform} class injection: flip 5\% of the labels in each class; (2) \textit{majority} class injection: flip 5\% of the labels in the majority class; and (3) \textit{minority} class injection: flip 5\% of the labels in the minority class. 
\begin{itemize}[noitemsep]
\item \textit{\textbf{cleanlab:}} We employ \textit{cleanlab}~\cite{northcutt2017cleanlab} to automatically clean the mislabels and then run the ML models, because it is model agnostic and can be easily configurable with any downstream model. \textit{cleanlab} implements confident learning with provable guarantees of exact noise estimation and label error finding.
\end{itemize}

\vspace{-2mm}
\subsection{Datasets}
\label{sec:cleanml_datasets}
We collected 14 real-world datasets\footnote{\scriptsize All datasets, their descriptions and error prevalence in each dataset can be found in \url{https://github.com/chu-data-lab/CleanML/blob/master/DatasetDescriptions.pdf}.} with varying error types and error rates, as summarized in Table \ref{T:dataset}.   

\vspace{-3mm}
\begin{table}[!h]
\caption{\scriptsize Dataset and Error Types}
\vspace{-2mm}
\centering
\resizebox{0.9\linewidth}{!}{
\begin{tabular}{c|c|c|c|c|c}
\toprule
\multirow{2}{*}{\textbf{Datasets}} & \multicolumn{5}{c}{\textbf{Error Types}}\\ \cline{2-6} 
& \multicolumn{1}{l|}{\textbf{Inconsistencies}} & \multicolumn{1}{l|}{\textbf{Duplicates}} & \multicolumn{1}{l|}{\textbf{Missing Values}} & \multicolumn{1}{l|}{\textbf{Outliers}} & \begin{tabular}[c]{@{}c@{}}\textbf{Mislabels} \end{tabular} \\ 
\midrule
Citation & & x & & & \\ 
EEG  & & &  & x &x \\ 
Marketing & & & x& & x  \\ 
Movie & x & x & & & \\ 
Company & x & & & & \\ 
Restaurant & x & x & & & \\ 
Sensor & & & & x & \\ 
Titanic & & & x &  & x\\ 
Credit & & &x  & x & \\ 
University & x & & & &\\ 
USCensus & & & x& & x\\ 
Airbnb & & x & x & x &  \\ 
BabyProduct & & & x & & \\ 
Clothing & & & &  & x \\ 
\bottomrule
\end{tabular}}
\label{T:dataset}
\vspace{-4mm}
\end{table}

\subsection{ML Models}
\label{sec:cleanml_models}
We selected seven classical and competitive ML models commonly used in classification tasks on structured datasets, including Logistic Regression, k-Nearest Neighbors (KNN), Decision Tree, Random Forest, AdaBoost, Naive Bayes and XGBoost~\cite{chen2016xgboost}. We use scikit-learn \cite{scikit-learn} for training models.

\subsection{Scenarios}
\label{sec:cleanml_scenarios}


Data cleaning may be applied either in the model development phase on the training data, or in the model deployment phase on the test data (c.f. Figure~\ref{fig:cleanml_workflow}). 
Depending on where data cleaning is applied, we can have four different performance metrics, as shown in~Table~\ref{approach}: (1) \textit{Case A} represents a model built using the original dirty training set and tested on the original dirty test set; (2) \textit{Case B} represents a model built using the original dirty training set and tested on the cleaned test set; (3) \textit{Case C} represents a model built using the cleaned training set and tested on the original dirty test set; and (4) \textit{Case D} represents a model built using the cleaned training set and tested on the cleaned test set. 

\begin{table}[!h]
\vspace{-9mm}
\centering
    \begin{subtable}{}
    \centering
\caption{\scriptsize Four different performance metrics on where data cleaning is performed}
\resizebox{.6\linewidth}{!}{
\begin{tabular}{ccc} 
\toprule
 & \textbf{Dirty Test Set}& \textbf{Cleaned Test Set}\\ 
 \midrule
\textbf{Dirty Training Set} & \textit{A} & \textit{B} \\ 
\textbf{Cleaned Training Set} & \textit{C} & \textit{D} \\ 
\bottomrule
\end{tabular}}
\label{approach}
    \end{subtable}
\vspace{-2mm}
    \begin{subtable}{}
\centering
\caption{\scriptsize Four different performance metrics on how missing value is handled}
\resizebox{.8\linewidth}{!}{
\begin{tabular}{ccc} 
\toprule
 &\textbf{Test Set by Deletion}&  \textbf{Test Set by Imputation}\\ 
 \midrule
\textbf{Training Set by Deletion} & \textit{A} & \textit{B} \\ 
\textbf{Training Set by Imputation}& \textit{C} & \textit{D} \\ \bottomrule
\end{tabular}}
\label{approach2}
    \end{subtable}
\end{table}

\vspace{-3mm}
We aim to evaluate the impact of cleaning on ML model performance both in the model development phase and in the model deployment phase using these four metrics: 

\stitle{Model Development Scenario (BD).} During the model development phase, ML developers would like to assess whether cleaning the training data can potentially improve the model performance on unseen test data. Therefore, we need to train two models: one trained on the original dirty training set and one trained on the cleaned version. To compare the two models on equal grounds, we need to evaluate them on the same test set. We can either compare the two models on the same dirty test set (i.e., compare \textit{Case A} with \textit{Case C}) or the same cleaned test set (i.e., compare \textit{Case B} with \textit{Case D}). We choose to compare \textit{Case B} with \textit{Case D} as it is common practice to ensure that the test set is cleaned for performance evaluations.

\stitle{Model Deployment Scenario (CD).} During the model deployment phase, a particular ML model has been trained and deployed into production to make predictions on incoming test data. In this phase, ML developers would like to know whether cleaning the test data can improve the performance of an already trained model. Hence, we can either compare \textit{Case A} with \textit{Case B}, or compare \textit{Case C} with \textit{Case D}. We choose to compare \textit{Case C} with \textit{Case D} because they use the cleaned training data, which often produces a better model.

\stitle{Special Consideration for Missing Values.} Missing values need special attention, as we cannot train a model or make predictions on a dataset with missing values. This means that \textit{Cases A}, \textit{B}, \textit{C} in Table~\ref{approach} are not available when dealing with missing values. 
To deal with this issue, we treat the dataset by deleting records with missing values as the ``dirty'' dataset, and a cleaned dataset is only achieved by filling in the missing values, as shown in~Table~\ref{approach2}.
In addition, we can only consider the model development scenario (BD) for missing values since deleting records from test set (i.e., \textit{Case A} and \textit{Case C}) is usually not acceptable in practice --- the missing values in test set have to be filled in to make predictions.  



%% file: cleanml_db_instance.tex
\section{CleanML Database Instance}
\label{sec:cleanml_instance}
\vspace{-2mm}
\begin{table}[!h]
\caption{\scriptsize Example of Experiment Specifications}
\vspace{-3mm}
\scalebox{0.65}{
\begin{tabular}{cccccc}
\multicolumn{6}{l}{$s_1$} \\ \hline
\multicolumn{1}{|c|}{\textbf{Dataset}} & \multicolumn{1}{c|}{\textbf{Error Type}} & \multicolumn{1}{c|}{\textbf{Detection}} & \multicolumn{1}{c|}{\textbf{Repair}} & \multicolumn{1}{c|}{\textbf{ML Model}} & \multicolumn{1}{c|}{\textbf{Scenario}} \\ \hline
\multicolumn{1}{|c|}{EEG} & \multicolumn{1}{c|}{Outliers} & \multicolumn{1}{c|}{IQR} & \multicolumn{1}{c|}{Mean Imputation} & \multicolumn{1}{c|}{Logistic Regression} & \multicolumn{1}{c|}{BD} \\ \hline
\end{tabular}}

\scalebox{0.72}{
\begin{tabular}{ccccc}
\multicolumn{1}{l}{$s_2$} & \multicolumn{1}{l}{} & \multicolumn{1}{l}{} & \multicolumn{1}{l}{} & \multicolumn{1}{l}{} \\ \hline
\multicolumn{1}{|c|}{\textbf{Dataset}} & \multicolumn{1}{c|}{\textbf{Error Type}} & \multicolumn{1}{c|}{\textbf{Detection}} & \multicolumn{1}{c|}{\textbf{Repair}} & \multicolumn{1}{c|}{\textbf{Scenario}} \\ \hline
\multicolumn{1}{|c|}{EEG} & \multicolumn{1}{c|}{Outliers} & \multicolumn{1}{c|}{IQR} & \multicolumn{1}{c|}{Mean Imputation} & \multicolumn{1}{c|}{BD} \\ \hline
\end{tabular}
}

\scalebox{0.75}{
\begin{tabular}{ccc}
\multicolumn{1}{l}{$s_3$} & \multicolumn{1}{l}{} & \multicolumn{1}{l}{} \\ \hline
\multicolumn{1}{|c|}{\textbf{Dataset}} & \multicolumn{1}{c|}{\textbf{Error Type}} & \multicolumn{1}{c|}{\textbf{Scenario}} \\ \hline
\multicolumn{1}{|c|}{EEG} & \multicolumn{1}{c|}{Outliers} & \multicolumn{1}{c|}{BD} \\ \hline
\end{tabular}
}
\label{example_outlier}
\vspace{-4mm}
\end{table}

In this section, we show how to populate the CleanML database instance. We have presented all possible values for all attributes in the primary keys of $\{R1, R2, R3\}$. Each combination of value assignment for the key in $R \in \{R1, R2, R3\}$ defines a particular setting, which we call \textit{experiment specification}. Our goal is to determine the value for the ``flag'' attribute for each experiment specification, i.e., what is the impact of cleaning on ML for a given experiment specification. 
Table~\ref{example_outlier} shows three example experiment specifications $s_1$, $s_2$ and $s_3$ in $R1$, $R2$ and $R3$, respectively.

\vspace{-2mm}
\subsection{Generating One Performance Metric Pair}
\label{sec:generate_one_pair}
Given an experiment specification $s_1$ in $R1$, we can generate a pair of performance metrics through following steps:
\begin{enumerate}[noitemsep, leftmargin=*, label={(\arabic*)}]
    \item \textbf{Splitting dataset.} We split the dataset in $s_1$ randomly into a training and a test set with a 70/30 ratio.
    \item \textbf{Performing data cleaning.} We clean the error type given in $s_1$ in both the training set and the test set, using the cleaning algorithm defined in $s_1$. To avoid any data leakage, all statistics necessary for data cleaning, such as mean, are computed only on the training set, and are used to clean both the training and the test set. 
    \item \textbf{Training ML models.} If the scenario is BD in $s_1$, we train two ML models, one on the original dirty training set and one on cleaned training set. If the scenario is CD, we only need to train one ML model on the cleaned training set. We perform hyper-parameter tunings using standard random search and 5-fold cross validation.
    \item \textbf{Evaluating ML models.} If the scenario is BD in $s_1$, we evaluate the two ML models on the cleaned test set to get a pair of performance metrics. If the scenario is CD, we evaluate the one model on the dirty test set and the clean test set, respectively, to get a pair of performance metrics. We use the classification accuracy as the metric on all datasets, except for the class-imbalanced datasets (e.g. Credit) for which we use F1 score as the metric.
\end{enumerate}

Given an experiment specification $s_2$ in $R2$ without a specific model, we perform the same four steps with the following modification for Step (3). We train all seven ML models (c.f. Section~\ref{sec:cleanml_models}) and select the model with the best validation accuracy from cross validation. The selected model is then used to obtain a pair of metrics in Step (4). 

Given an experiment specification $s_3$ in $R3$ without a specific model and a specific cleaning method, we try all available cleaning methods (c.f. Section~\ref{sec:cleanml_cleaning_attributes}) for the error type given in $s_3$ to clean dataset at Step (2). We will select the cleaning method and the ML model that has the best validation accuracy in Step (3). The selected model and cleaning method is then used to obtain a pair of metrics in Step (4).


\vspace{-8mm}
\begin{table}[!h]
\centering
\begin{subtable}{}
\centering
\caption{\scriptsize $s_1$ Metric Pairs}
\label{example_outliers_r1_one}
\scalebox{0.78}{
\begin{tabular}{ccccc}
\hline
\multicolumn{1}{c|}{\multirow{2}{*}{\textbf{Model}}} & \multicolumn{2}{c|}{\textit{Train on Dirty Training set}} & \multicolumn{2}{c}{\textit{Train on Clean Training set}} \\ \cline{2-5} 
\multicolumn{1}{c|}{} & \multicolumn{1}{c|}{\textbf{Val Acc}} & \multicolumn{1}{c|}{\textbf{Clean Test Acc}} & \multicolumn{1}{c|}{\textbf{Val Acc}} & \textbf{Clean Test Acc} \\ \hline
\multicolumn{1}{c|}{LR} & \multicolumn{1}{c|}{0.638} & \multicolumn{1}{c|}{0.634} & \multicolumn{1}{c|}{0.673} & 0.668 \\ \hline
\multicolumn{5}{c}{Metric Pair: (0.634, 0.668)} \\ \hline
\end{tabular}
}
\end{subtable}
\begin{subtable}{}
\centering
\caption{\scriptsize $s_2$ Metric Pairs}
\label{example_outliers_r2_one}
\scalebox{0.78}{
\begin{tabular}{ccccc}
\hline
\multicolumn{1}{c|}{\multirow{2}{*}{\textbf{Model}}} & \multicolumn{2}{c|}{\textit{Train on Dirty Training set}} & \multicolumn{2}{c}{\textit{Train on Clean Training set}} \\ \cline{2-5} 
\multicolumn{1}{c|}{} & \multicolumn{1}{c|}{\textbf{Val Acc}} & \multicolumn{1}{c|}{\textbf{Clean Test Acc}} & \multicolumn{1}{c|}{\textbf{Val Acc}} & \textbf{Clean Test Acc} \\ \hline
\multicolumn{1}{c|}{Adaboost} & \multicolumn{1}{c|}{0.763} & \multicolumn{1}{c|}{0.711} & \multicolumn{1}{c|}{0.718} & 0.715 \\
\multicolumn{1}{c|}{KNN} & \multicolumn{1}{c|}{0.895} & \multicolumn{1}{c|}{0.821} & \multicolumn{1}{c|}{\textbf{0.948}} & \textbf{0.956} \\
\multicolumn{1}{c|}{XGBoost} & \multicolumn{1}{c|}{\textbf{0.932}} & \multicolumn{1}{c|}{\textbf{0.862}} & \multicolumn{1}{c|}{0.920} & 0.922 \\ \hline
\multicolumn{5}{c}{Metric Pair: (0.862, 0.956)} \\ \hline
\end{tabular}
}
\end{subtable}
\begin{subtable}{}
\centering
\caption{\scriptsize $s_3$ Metric Pairs}
\label{example_outliers_r3_one}
\scalebox{0.8}{
\begin{tabular}{ccccc}
\hline
\multicolumn{1}{c|}{\multirow{2}{*}{\textbf{Detection}}} & \multicolumn{1}{c|}{\multirow{2}{*}{\textbf{Repair}}} & \multicolumn{1}{c|}{\textit{\begin{tabular}[c]{@{}c@{}}Best Model on \\ Clean Training Set\end{tabular}}} & \multicolumn{1}{c|}{\textit{\begin{tabular}[c]{@{}c@{}}Best Model on\\ Dirty Training Set\end{tabular}}} & \textit{\begin{tabular}[c]{@{}c@{}}Best Model on \\ Clean Training Set\end{tabular}} \\ \cline{3-5} 
\multicolumn{1}{c|}{} & \multicolumn{1}{c|}{} & \multicolumn{1}{c|}{\textbf{Val Acc}} & \multicolumn{1}{c|}{\textbf{Clean Test Acc}} & \textbf{Clean Test Acc} \\ \hline
\multicolumn{1}{c|}{\textbf{SD}} & \multicolumn{1}{c|}{\textbf{Mean}} & \multicolumn{1}{c|}{\textbf{0.959}} & \multicolumn{1}{c|}{\textbf{0.937}} & \textbf{0.969} \\
\multicolumn{1}{c|}{SD} & \multicolumn{1}{c|}{Median} & \multicolumn{1}{c|}{0.955} & \multicolumn{1}{c|}{0.938} & 0.964 \\
\multicolumn{1}{c|}{SD} & \multicolumn{1}{c|}{Mode} & \multicolumn{1}{c|}{0.955} & \multicolumn{1}{c|}{0.937} & 0.964 \\
\multicolumn{1}{c|}{...} & \multicolumn{1}{c|}{...} & \multicolumn{1}{c|}{...} & \multicolumn{1}{c|}{...} & ... \\ \hline
\multicolumn{5}{c}{Metric Pair: (0.937, 0.969)} \\ \hline
\end{tabular}
}
\end{subtable}
\begin{subtable}{}
\centering
\caption{\scriptsize Accuracy Evaluated on the Clean Test Set}
\label{results_acc_clean}
\scalebox{0.78}{
\begin{tabular}{c|ccccccc}
\hline
\textbf{Split Seed} & 1 & 2 & 3 & 4 & ... & 19 & 20 \\ \hline
\textbf{B} & 0.632 & 0.631 & 0.634 & 0.638 & ... & 0.629 & 0.632 \\
\textbf{D} & 0.657 & 0.674 & 0.668 & 0.676 & ... & 0.669 & 0.668 \\ \hline
\end{tabular}
}
\end{subtable}
\end{table}
\vspace{-2mm}
\begin{exmp} 
To generate \textit{a metric pair for $s_1$} in Table \ref{example_outlier}, we first split EEG into training/test sets. We detect outliers in the training set and test set using IQR detection and repair them with mean imputation. The quantiles used in detection and mean used in repair are computed on the training set. Since the scenario here is BD, we train two logistic regression models on the dirty training set and the cleaned training set, respectively. Finally, we evaluate the two models on the cleaned test set to obtain two test accuracy scores to form a metric pair, as shown in Table~\ref{example_outliers_r1_one}. To generate \textit{a metric pair for $s_2$}, we train all seven ML models. As shown in Table~\ref{example_outliers_r2_one}, based on the validation accuracy, XGBoost is the best model trained on the dirty training set, and KNN is the best model on the cleaned training set. We then evaluate the two  models on the same cleaned test set to get two accuracy scores that form a metric pair. To generate \textit{a metric pair for $s_3$}, in addition to model selection, we use all available methods for cleaning outliers. As shown in Table~\ref{example_outliers_r3_one}, detecting outliers by SD and repairing by mean imputation is the best cleaning method, as it has the best validation accuracy among all cleaning methods. Hence, we use its metric pair as the metric pair for $s_3$.

\end{exmp}
\vspace{-2mm}
\subsection{Handling Randomness}
\label{sec:control_randomness}

While the procedure described in Section~\ref{sec:generate_one_pair} produces one pair of metrics given an experiment specification, it may produce an entirely different pair of metrics for a different train/test split.  To handle ML randomness, we conduct the same procedure 20 times, each time with a different train/test split. Table~\ref{results_acc_clean} shows 20 pairs of metrics we obtained given $s_1$ in Table~\ref{example_outlier}. 
Given the 20 metric pairs, one might be tempted to take the average and set the ``flag'' attribute based on the difference between the two metrics in the averaged metric pair: if the averaged $D > B$, flag is set to ``P'' (positive); if the averaged $D < B$, flag is set to ``N'' (negative); if the averaged $D = B$, flag is set to ``S'' (insignificant);
This simple approach is problematic since the absolute difference between the two averaged metrics may not be significant enough. 

We propose to follow a rigorous statistical significance testing procedure to determine the flag based on 20 metric pairs. Specifically, we use the \textit{paired sample $t$-test}~\cite{mcdonald2009handbook}, which is commonly used to determine whether the mean difference between two sets of observations is zero, positive, or negative. For example, it can be used to determine whether taking a course has a positive impact on student academic performance by comparing test scores of 20 students before and after taking the course. 
As far as we know, existing applications of paired sample t-test are often used to determine a binary outcome (e.g., positive or non-positive). However, our application needs to produce a three-valued flag attribute. We describe our process to determine the flag attribute in the following.

Let $\mu^t$ be the mean difference of the performance metrics before and after data cleaning. We use three paired sample t-test with the following null and alternative hypotheses:
\begin{table}[!h]
\centering
\vspace{-2mm}
\scalebox{0.78}{
\begin{tabular}{c|c|c|c}
\toprule
\textbf{Hypothesis} & \textbf{Two-tailed $t$-test} & \textbf{Upper-tailed $t$-test} & \textbf{Lower-tailed $t$-test }\\
\midrule
Null & $H_{0}^t: \mu^t = 0$ & $H_1^t: \mu^t \leq 0$ & $H_2^t: \mu^t \geq 0$ \\ 
Alternative & $H_a^t: \mu^t \neq 0$ & $H_b^t: \mu^t > 0$ & $H_c^t: \mu^t < 0$ \\ 
\bottomrule
\end{tabular}
}
\vspace{-3mm}
\end{table}


 



We determine the flag attribute based on the three tests as follows, where  $p_0$, $p_1$, $p_2$ denote the $p$-values of two-tailed $t$-test, upper-tailed $t$-test, and lower-tailed $t$-test respectively and $\alpha$ denotes the significant level.  
\begin{enumerate}[noitemsep,label={(\arabic*)},font=\scriptsize]
    \item \scriptsize if $p_0 \geq \alpha$, Flag = ``S".
    \item if $p_0 < \alpha \text{ and } p_1 < \alpha$, Flag = ``P".
    \item if $p_0 < \alpha \text{ and } p_2 < \alpha$, Flag = ``N".
\end{enumerate}

The intricacy of conducting three tests together lies in the fact that, if the test statistics distribution is symmetric (e.g., Gaussian), the $p$-value in one of the one-tailed tests is exactly half of the $p$-value in the two-tailed test. Hence, a two-tailed test with significance implies that one of the one-tailed tests is significant; yet if the one-tailed test is significant, the two-tailed one is not necessarily significant. What is criticized often by people is to only report the significance by a one-tailed test because the two-tailed test is insignificant. However, we do not face this claim because we conduct three tests and only report the one-tailed test results if the two-tailed test is significant. 



\begin{exmp}For $s_1$, we run these three tests on 20 metric pairs (c.f. Table~\ref{results_acc_clean}). The three p-values we obtain are: $p_0 = 3.82e^{-17}$, $p_1 = 1.91e^{-17}$, and $p_2 = 1$. $\alpha$ is set to $0.05$. Hence, we determine the ``flag'' attribute for $s_1$ as ``P''. 


\end{exmp}



\vspace{-1mm}
\subsection{Controlling False Discoveries}
\label{sec:control_fdr}



Given the comprehensive scope we have in our study, we have many hypothesis testing procedures to run, which can cause \textit{false discoveries}, namely, many hypotheses are incorrectly declared as significant.
This is commonly known as the \textit{multiple hypothesis testing} problem in the statistics literature~\cite{rupert2012simultaneous}. To see the effect of multiple testing, consider a case where there are $20$ hypotheses to test and we set a significance level of $\alpha = 0.05$. The probability of observing at least one significant result due to chance is $1 - (1-\alpha)^{20} \approx 0.64$. Thus, we have at least $64\%$ chance of observing significant results within just 20 tests, even if all tests are actually insignificant. 

With 3612, 516 and 168 hypotheses in our relations $R1$, $R2$ and $R3$ (3 $\times$ the number of unique assignments for key attributes), respectively, it is highly likely that our results contain many false discoveries by chance. Strategies to control false discovery rate caused by the multiple hypothesis testing problem usually adjust the significance level $\alpha$ in some way~\cite{rupert2012simultaneous,benjamini1995controlling}. E.g., a simple way to adjust $\alpha$ is called the \textit{Bonferroni correction}~\cite{Bonferroni36}, which uses $\frac{\alpha}{m}$ instead of $\alpha$, where $m$ is the number of tests. However, this correction can fail with the presentation of more tests with non-significant results. 

Instead of adjusting the significance level $\alpha$ for every test, another strategy is to rank the tests by their \textit{$p$-values} in an ascending order, and then select a certain number of top ranked tests as significant. This is called the FDR approach~\cite{friedman2001elements}, which ensures that in expectation the false discovery rate 
is below a user-defined threshold $\alpha$.
Common FDR approaches are Benjamini-Hochberg (BH) and Benjamini-Yekutieli (BY) procedures \cite{friedman2001elements}, which differ in how many top ranked tests to declare as significant. 
We employ the BY procedure since it controls the FDR under arbitrary dependence assumptions, which is appropriate for this study because two tests should not be considered independent if their experiment specifications have common attributes. 
For each relation in $\{R1, R2, R3\}$, we conduct a separate BY procedure and use $\alpha = 0.05$.

%% file: cleanml_db_analysis.tex
\section{Analyzing CleanML Database}
\label{sec:cleanml_analysis}

We first present our strategy for performing result analysis in Section~\ref{analysis_method}, and discuss the detailed findings for each error type in Sections~\ref{sec:analysis_mv} to~\ref{sec:analysis_dup}.


\vspace{-2mm}
\subsection{Strategy for Result Analysis}\label{analysis_method}

We investigate the impact of data cleaning on ML by running SQL queries on $R \in \{R1, R2, R3\}$ (c.f. Table~\ref{dcml_schema}). We first present the SQL query templates, where $E \in$ \{inconsistencies, duplicates, mislabels, outliers, missing\ values\}, and then discuss two angles for analyzing the results of SQL queries. 


\stitle{Varying Granularity of Analysis.} The impact of cleaning on ML depends on a variety of factors, warranting varying granularities of analysis.
We first fix the error type and group by the flag attribute (Q1), which aggregates over all datasets, ML models, scenarios, and cleaning methods. This gives the impact of cleaning the given type of error in general. 
Then we group by an additional attribute to see if there is any scenario (Q2), ML model (Q3), cleaning method (Q4.1 for detection and Q4.2 for repair), or dataset (Q5), where the  impact is different from the general trend we observe from Q1.

\stitle{Compare $R1$, $R2$, and $R3$.} We also investigate the difference of results between the same query template issued against different relations. This indicates whether performing model selection (i.e., $R2$) and cleaning algorithm selection (i.e., $R3$) are helpful for achieving a positive impact when applying data cleaning on downstream ML models.  
Due to space limitation, throughout the remaining analysis, we will only show  the results of those SQL queries issued against those relations that convey nontrivial and unique findings. We refer readers to the technical report~\cite{cleanml_report} for all query results.
\vspace{-3mm}

\begin{figure}[!h]
    \centering
    \includegraphics[width=\columnwidth]{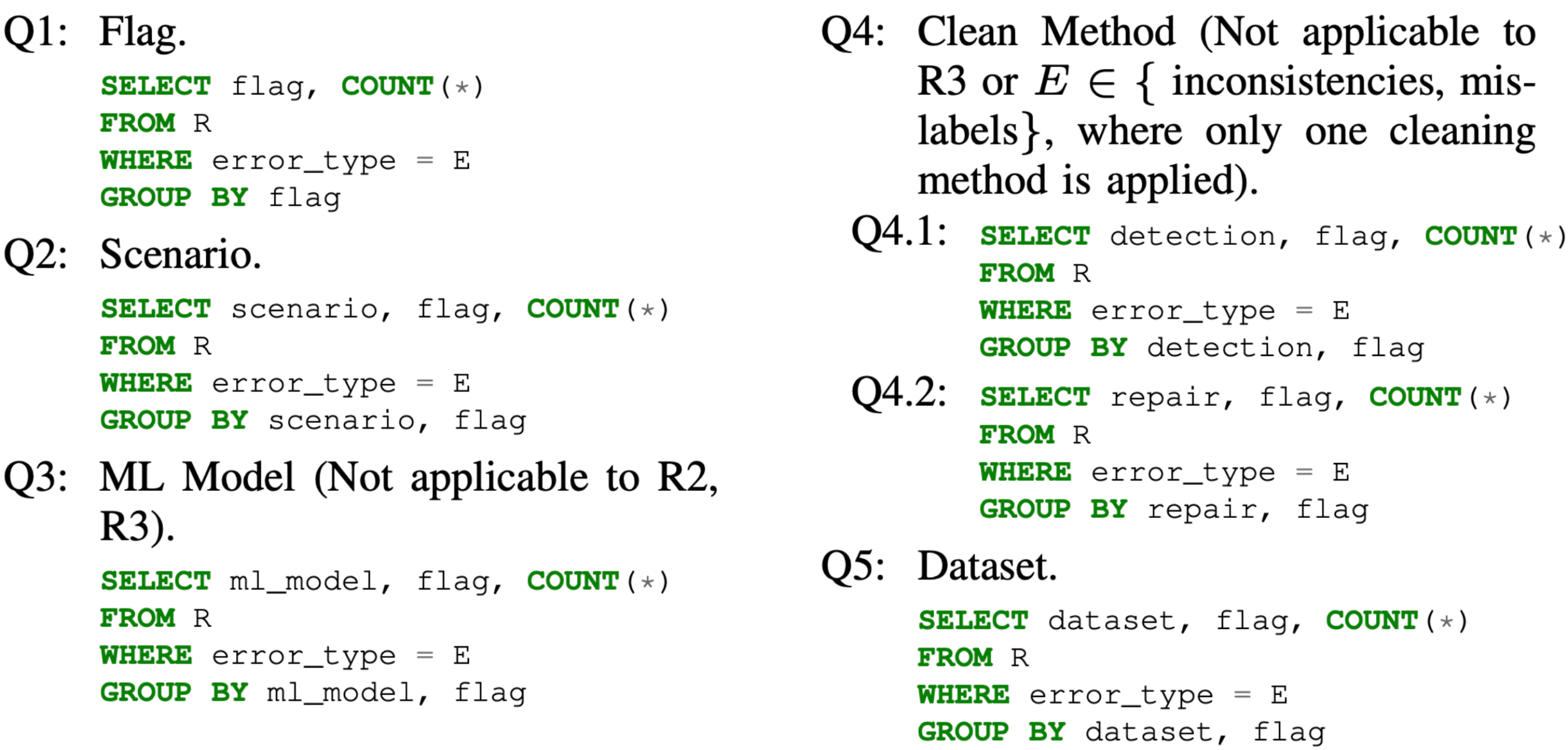}
    \label{fig:sql_code}
\end{figure}


\subsection{Missing Values}
\label{sec:analysis_mv}

\begin{table}[!h]
\caption{Query Results for Missing Values}
\vspace{-2mm}
\centering
\scalebox{0.78}{
\begin{tabular}{c|c|c|c|c}
\multicolumn{5}{c}{\textbf{Q1 (E = Missing Values)}} \\
\toprule
\multicolumn{2}{c|}{\textbf{R}} & \textbf{P} & \textbf{S} & \textbf{N} \\
\midrule
\multicolumn{2}{c|}{R1} & 49\% (143) & 27\% (80) & 24\% (71) \\
\multicolumn{2}{c|}{R2} & 57\% (24) & 21\% (9) & 21\% (9) \\
\multicolumn{2}{c|}{R3} & 33\% (2) & 50\% (3) & 17\% (1) \\
\bottomrule
\multicolumn{5}{c}{\textbf{Q4.2 (E = Missing Values)}} \\
\toprule
\textbf{R} & \textbf{Imputation} & \textbf{P} & \textbf{S} & \textbf{N} \\
\midrule
\multirow{7}{*}{R1} & HoloClean & 38\% (16) & 29\% (12) & 33\% (14) \\
 & Mean\_Dummy & 52\% (22) & 24\% (10) & 24\% (10) \\
 & Mean\_Mode & 50\% (21) & 29\% (12) & 21\% (9) \\
 & Median\_Dummy & 52\% (22) & 24\% (10) & 24\% (10) \\
 & Median\_Mode & 64\% (27) & 31\% (13) & 5\% (2) \\
 & Mode\_Dummy & 52\% (22) & 24\% (10) & 24\% (10) \\
 & Mode\_Mode & 31\% (13) & 31\% (13) & 38\% (16) \\
 \midrule
\multirow{7}{*}{R2} & HoloClean & 50\% (3) & 17\% (1) & 33\% (2) \\
 & Mean\_Dummy & 67\% (4) & 17\% (1) & 17\% (1) \\
 & Mean\_Mode & 67\% (4) & 17\% (1) & 17\% (1) \\
 & Median\_Dummy & 50\% (3) & 33\% (2) & 17\% (1) \\
 & Median\_Mode & 83\% (5) & 17\% (1) & 0\% (0) \\
 & Mode\_Dummy & 50\% (3) & 33\% (2) & 17\% (1) \\
 & Mode\_Mode & 33\% (2) & 17\% (1) & 50\% (3) \\
 \bottomrule
\multicolumn{5}{c}{\textbf{Q5 (E = Missing Values)}} \\
\toprule
\textbf{R} & \textbf{Dataset} & \textbf{P} & \textbf{S} & \textbf{N} \\
\midrule
\multirow{6}{*}{R1} & Airbnb & 6\% (3) & 88\% (43) & 6\% (3) \\
 & BabyProduct & 57\% (28) & 0\% (0) & 43\% (21) \\
 & Credit & 29\% (14) & 67\% (33) & 4\% (2) \\
 & Marketing & 49\% (24) & 8\% (4) & 43\% (21) \\
 & Titanic & 65\% (32) & 0\% (0) & 35\% (17) \\
 & USCensus & 86\% (42) & 0\% (0) & 14\% (7) \\
 \bottomrule
\end{tabular}
}
\label{table:result_missing_values}
\vspace{-4mm}
\end{table}
Table \ref{table:result_missing_values} shows the results of SQL queries for missing values issued against all three relations.
\textbf{(Q1):} As we can see in the results of Q1, cleaning missing values by imputation mostly improves the performance or achieves similar performance as deleting records with missing values. However, there are still quite a few ``N" flags in the results of $R1$ and $R2$, which indicates that imputation may be worse than deletion. This is because values filled in by imputation algorithms are various ``guesses'', which may be far off the ground-truth values.
Comparing the results of Q1 on $R1$, $R2$, we observe that imputation is more likely to have positive impact and less likely to have negative impact on better ML models. Comparing the results of Q1 on $R2$ and $R3$, we can see that selecting the best imputation method using a validation set can further reduce the negative impacts. 
\textbf{(Q2):} As mentioned in Section \ref{sec:cleanml_scenarios}, we only consider the scenario BD for missing values,  hence Q2 is not applicable.
\textbf{(Q3):} All models share similar results as we observe in Q1; a result table hence is omitted here.
\textbf{(Q4):} Since there is only one detection method for missing values, Q4.1 is not applicable. From the results of Q4.2 on R1 and R2 (Q4.2 is not applicable to R3), we can see that there is no significant difference between impacts of different imputation methods. In particular, the state-of-the-art probabilistic imputation method HoloClean~\cite{rekatsinas2017holoclean} has no clear advantage over other simple imputation methods.
\textbf{(Q5):} We show the results of Q5 issued against R1 only, as the results of Q5 issued against R2 and R3 reveal similar findings. 
%
As we can see, the impact of data cleaning varies significantly across different datasets. 
For example, the ``USCensus" dataset has more ``P"\%, while ``Airbnb" dataset has more ``S"\%. 

\stitle{Summary of Observations for Missing Values.} (1) Cleaning  missing  values  by imputation is more likely to improve ML or achieve similar performance compared with deletion; (2) with model and imputation method selection, we are more likely to observe positive impacts and less likely to observe negative impacts; (3) impacts vary largely across different datasets and advanced cleaning methods such as  HoloClean~\cite{rekatsinas2017holoclean} is not noticeably better than simple imputation methods. 


\vspace{-2mm}
\subsection{Outliers}
\vspace{-2mm}
\label{sec:analysis_outliers}

\begin{table}[!h]
\centering
\caption{Query Results for Outliers}
\vspace{-2mm}
\scalebox{0.78}{
\begin{tabular}{c|c|c|c|c}
\multicolumn{5}{c}{\textbf{Q1 (E = Outliers)}} \\
\toprule
\multicolumn{2}{c|}{\textbf{R}} & \textbf{P} & \textbf{S} & \textbf{N} \\
\midrule
\multicolumn{2}{c|}{R1} & 31\% (176) & 61\% (339) & 8\% (45) \\
\multicolumn{2}{c|}{R2} & 39\% (31) & 56\% (45) & 5\% (4) \\
\multicolumn{2}{c|}{R3} & 12\% (1) & 88\% (7) & 0\% (0) \\
\bottomrule
\multicolumn{5}{c}{\textbf{Q3 (E = Outliers)}} \\
\toprule
\textbf{R} & \textbf{Model} & \textbf{P} & \textbf{S} & \textbf{N} \\
\midrule
\multirow{7}{*}{R1} & Adaboost & 12\% (10) & 70\% (56) & 18\% (14) \\
 & Decision Tree & 30\% (24) & 69\% (55) & 1\% (1) \\
 & Gussian Naive Bayes & 31\% (25) & 64\% (51) & 5\% (4) \\
 & KNN & 52\% (42) & 42\% (34) & 5\% (4) \\
 & Logistic Regression & 22\% (18) & 60\% (48) & 18\% (14) \\
 & Random Forest & 32\% (26) & 60\% (48) & 8\% (6) \\
 & XGboost & 39\% (31) & 59\% (47) & 2\% (2) \\
 \bottomrule
\multicolumn{5}{c}{\textbf{Q4.1 (E = Outliers)}} \\
\toprule
\textbf{R} & \textbf{Detect} & \textbf{P} & \textbf{S} & \textbf{N} \\
\midrule
\multirow{3}{*}{R1} & IF & 34\% (57) & 47\% (79) & 19\% (32) \\
 & IQR & 59\% (99) & 38\% (64) & 3\% (5) \\
 & SD & 8\% (13) & 90\% (151) & 2\% (4) \\
 \midrule
\multirow{3}{*}{R2} & IF & 38\% (9) & 58\% (14) & 4\% (1) \\
 & IQR & 71\% (17) & 17\% (4) & 12\% (3) \\
 & SD & 17\% (4) & 83\% (20) & 0\% (0) \\
 \bottomrule
\multicolumn{5}{c}{\textbf{Q4.2 (E = Outliers)}} \\
\toprule
\textbf{R} & \textbf{Repair} & \textbf{P} & \textbf{S} & \textbf{N} \\
\midrule
\multirow{4}{*}{R1} & HoloClean & 12\% (7) & 80\% (45) & 7\% (4) \\
 & Mean & 33\% (56) & 60\% (101) & 7\% (11) \\
 & Median & 33\% (56) & 58\% (97) & 9\% (15) \\
 & Mode & 34\% (57) & 57\% (96) & 9\% (15) \\
 \midrule
\multirow{4}{*}{R2} & HoloClean & 12\% (1) & 88\% (7) & 0\% (0) \\
 & Mean & 42\% (10) & 54\% (13) & 4\% (1) \\
 & Median & 46\% (11) & 50\% (12) & 4\% (1) \\
 & Mode & 38\% (9) & 54\% (13) & 8\% (2) \\
 \bottomrule
\multicolumn{5}{c}{\textbf{Q5 (E = Outliers)}} \\
\toprule
\textbf{R} & \textbf{Dataset} & \textbf{P} & \textbf{S} & \textbf{N} \\
\midrule
\multirow{4}{*}{R1} & Airbnb & 10\% (14) & 87\% (122) & 3\% (4) \\
 & Credit & 14\% (20) & 70\% (98) & 16\% (22) \\
 & EEG & 57\% (80) & 41\% (57) & 2\% (3) \\
 & Sensor & 44\% (62) & 44\% (62) & 11\% (16)\\
 \bottomrule
\end{tabular}
}
\label{table:result_outliers}
\vspace{-5mm}
\end{table}
Table \ref{table:result_outliers} shows the query results for outliers issued against all three relations.
\textbf{(Q1):} We can see that cleaning outliers mostly has an insignificant impact on model performance. This is in contrast to the findings for missing values, where cleaning mostly have a positive impact. We believe that it is because cleaning outliers is a fundamentally harder task --- while detecting missing values is trivial, detecting outliers is prone to mistakes, which may declare non-outliers as outliers and miss true outliers. 
Comparing the results of R1, R2 and R3 shows that with model selection and cleaning method selection, ``N"\% gradually decreases to 0. This indicates that using model and cleaning method selection can eliminate some negative impacts and improve robustness without losing much benefit, which is similar to the findings for missing values. 
\textbf{(Q2):} The findings of Q2 are similar to Q1, and hence are omitted.
\textbf{(Q3):} KNN has mostly ``P" flags, less ``N" and ``S" flags than most of the other ML models. It is because that KNN relies on the distances between examples to make predictions, and outliers can impact distances significantly. The differences between other models are not prominent. 
\textbf{(Q4.1):} As we see in Q4.1 on R1, IF and IQR have more ``P" flags and ``N" flags than the SD method. This indicates that IF and IQR are more aggressive than SD. Indeed, by examining the number of outliers detected by these methods, we learn that IF and IQR generally produce many more outliers than SD.
\textbf{(Q4.2): }The results of Q4.2 show that there is no significant difference between repair methods in R1 and R2. Similar to missing values, it suggests that advanced data cleaning methods such as HoloClean~\cite{rekatsinas2017holoclean} are not noticeable better than simple cleaning methods in terms of improving downstream ML models.
\textbf{(Q5):} We only show results of Q5 on R1, as the results on R2 and R3 show similar findings. We can see that most negative flags are from the ``Credit" dataset. In all relations, ``EEG" and ``Sensor" have more ``P" flags than other datasets. This echoes our observations for missing values: the impact of cleaning varies significantly across datasets. However, for outliers, we are not able to provide more detailed explanations as we do not even know whether the detected outliers are true outliers.

\stitle{Summary of Observations for Outliers.} (1) Cleaning outliers is more likely to have insignificant impact on model performance, at least for the set of outlier cleaning algorithms we considered; (2) with model selection and cleaning algorithm selection, the probability of having negative impacts can be greatly reduced; (3) the impact of cleaning varies vastly across  datasets; (4) since outliers are harder to clean, different detection methods have major differences in observed impact.
\begin{table}[t]
\centering
\caption{Query Results for Mislabel}
\vspace{-2mm}
\scalebox{0.78}{
\begin{tabular}{c|c|c|c|c}
\multicolumn{5}{c}{\textbf{Q1 (E = Mislabel)}} \\
\toprule
\multicolumn{2}{c|}{\textbf{R}} & \textbf{P} & \textbf{S} & \textbf{N} \\
\midrule
\multicolumn{2}{c|}{R1} & 47\% (85) & 38\% (70) & 15\% (27) \\
\multicolumn{2}{c|}{R2 \& R3} & 54\% (14) & 31\% (8) & 15\% (4) \\
\bottomrule
\multicolumn{5}{c}{\textbf{Q2 (E = Mislabel)}} \\
\toprule
\textbf{R} & \textbf{Scenario} & \textbf{P} & \textbf{S} & \textbf{N} \\
\midrule
\multirow{2}{*}{R1} & BD & 19\% (17) & 59\% (54) & 22\% (20) \\
 & CD & 75\% (68) & 18\% (16) & 8\% (7) \\
 \midrule
\multirow{2}{*}{R2 \& R3} & BD & 23\% (3) & 46\% (6) & 31\% (4) \\
 & CD & 85\% (11) & 15\% (2) & 0\% (0) \\
 \bottomrule
\multicolumn{5}{c}{\textbf{Q3 (E = Mislabel)}} \\
\toprule
\textbf{R} & \textbf{Model} & \textbf{P} & \textbf{S} & \textbf{N} \\
\midrule
\multirow{7}{*}{R1} & Adaboost & 54\% (14) & 31\% (8) & 15\% (4) \\
 & Decision Tree & 46\% (12) & 46\% (12) & 8\% (2) \\
 & Gussian Naive Bayes & 38\% (10) & 42\% (11) & 19\% (5) \\
 & KNN & 38\% (10) & 42\% (11) & 19\% (5) \\
 & Logistic Regression & 50\% (13) & 38\% (10) & 12\% (3) \\
 & Random Forest & 50\% (13) & 31\% (8) & 19\% (5) \\
 & XGboost & 50\% (13) & 38\% (10) & 12\% (3) \\
 \bottomrule
\multicolumn{5}{c}{\textbf{Q5 (E = Mislabel)}} \\
\toprule
\textbf{R} & \textbf{Dataset} & \textbf{P} & \textbf{S} & \textbf{N} \\
\midrule
\multirow{13}{*}{R1} & Clothing & 21\% (3) & 14\% (2) & 64\% (9) \\
 & EEG\_major & 43\% (6) & 29\% (4) & 29\% (4) \\
 & EEG\_minor & 50\% (7) & 29\% (4) & 21\% (3) \\
 & EEG\_uniform & 71\% (10) & 29\% (4) & 0\% (0) \\
 & Marketing\_major & 57\% (8) & 43\% (6) & 0\% (0) \\
 & Marketing\_minor & 86\% (12) & 14\% (2) & 0\% (0) \\
 & Marketing\_uniform & 64\% (9) & 36\% (5) & 0\% (0) \\
 & Titanic\_major & 21\% (3) & 79\% (11) & 0\% (0) \\
 & Titanic\_minor & 7\% (1) & 79\% (11) & 14\% (2) \\
 & Titanic\_uniform & 50\% (7) & 21\% (3) & 29\% (4) \\
 & USCensus\_major & 43\% (6) & 36\% (5) & 21\% (3) \\
 & USCensus\_minor & 43\% (6) & 50\% (7) & 7\% (1) \\
 & USCensus\_uniform & 50\% (7) & 43\% (6) & 7\% (1)\\
 \bottomrule
\end{tabular}
}
\vspace{-7mm}
\label{table:result_mislabel}
\end{table}

\vspace{-4mm}
\subsection{Mislabels}
\label{sec:mislabel_analysis}
Table \ref{table:result_mislabel} shows the query results for mislabels.
\textbf{(Q1):} We observe that cleaning mislabels mostly improves or insignificantly affects the performance of ML models. However, there are still a few ``N" flags in R1 and R2. This is because the automatic mislabel cleaning method cleanlab can make mistakes in detecting and correcting mislabels, which may harm the model performance. Comparing the results of R1 and R2, with model selection (R2), it becomes more likely to have positive impacts.
\textbf{(Q2):} We observe that in scenario BD, it is more likely to have insignificant impact, while it is more likely to have positive impact in scenario CD. This is because mislabels in the test set can directly change the accuracy (flipping labels can directly make true positives become false positives). However, mislabels in the training set affect the final test accuracy indirectly. Hence, the impact is less significant in BD than that in CD.
\textbf{(Q3):} 
We notice that AdaBoost and XGBoost are more reactive to mislabels, because boosting procedures assign higher weights to on those instances that are predicted incorrectly. Therefore, cleaning mislabeled examples largely alleviates negative impacts on boosting ML. 
\textbf{(Q4):} Q4 is not applicable for mislabels because we have only one way of cleaning, i.e., cleanlab.
\textbf{(Q5):} 
We observe that the impact varies vastly across datasets. For example, in ``Marketing\_major" dataset, there are more  ``P"\%, while in ``Clothing" dataset, there are more ``N"\%.  The results of Q5 issued against R2 and R3 (omitted here) reveal similar findings to that of Q1. 


\stitle{Summary of Observations for Mislabels:} (1) Cleaning mislabels is likely to have positive or insignificant impacts on ML; (2) With model selection, it is more likely to have positive impacts.
(3) boosting based ML models are most reactive to mislabels. (4) the impact varies significantly across datasets.
\vspace{-1mm}
\subsection{Inconsistencies}
\label{sec:analysis_incon}

\vspace{-2mm}
\begin{table}[!h]
\caption{Query Results for Inconsistencies}
\centering
\vspace{-2mm}
\scalebox{0.78}{
\begin{tabular}{c|c|c|c|c}

\multicolumn{5}{c}{\textbf{Q1 (E = Inconsistencies)}} \\
\toprule
\multicolumn{2}{c|}{\textbf{R}} & \textbf{P} & \textbf{S} & \textbf{N} \\
\midrule
\multicolumn{2}{c|}{R1} & 12\% (7) & 88\% (49) & 0\% (0) \\
\multicolumn{2}{c|}{R2 \& R3} & 25\% (2) & 75\% (6) & 0\% (0) \\
\bottomrule
\multicolumn{5}{c}{\textbf{Q5 (E = Inconsistencies)}} \\
\toprule
\textbf{R} & \textbf{Dataset} & \textbf{P} & \textbf{S} & \textbf{N} \\
\midrule
\multirow{4}{*}{R1} & Company & 29\% (4) & 71\% (10) & 0\% (0) \\
 & Movie & 14\% (2) & 86\% (12) & 0\% (0) \\
 & Restaurant & 0\% (0) & 100\% (14) & 0\% (0) \\
 & University & 7\% (1) & 93\% (13) & 0\% (0)\\
 \bottomrule
\end{tabular}
\vspace{-10mm}
}
\label{table:result_inconsistencies}
\end{table}
\vspace{-2mm}



Table \ref{table:result_inconsistencies} shows the query results for inconsistencies.
\textbf{(Q1):} The results of Q1 shows no negative impact when cleaning inconsistencies, though most experiments show insignificant impact.
Furthermore, comparing results of $R1$ with $R2$, selecting the best ML model increases the likelihood of observing positive impact  after cleaning inconsistencies.
\textbf{(Q2, Q3, Q4):} Grouping by the additional scenario attribute Q2 and  ML model attribute Q3 reveals similar findings as Q1. Since we only have one cleaning method for inconsistencies, Q4 is not applicable. Their results are hence omitted.
\textbf{(Q5): }From the results of Q5, in general, the pattern holds that insignificant impact of cleaning inconsistency prevails and no negative impacts of cleaning inconsistency is found. However,  distribution of ``P" flags and ``N" flags  varies significantly across datasets. For example, ``Company" and ``Movie" datasets have more ``P" flags due to the much greater number of inconsistencies in these two datasets upon examining the cleaning results.


\vspace{-1mm}
\stitle{Summary of Observations for Inconsistencies:} (1) Cleaning inconsistencies is more likely to have insignificant impact and unlikely to have negative impact on ML; (2) the impact varies significantly across datasets.


\vspace{-1mm}
\subsection{Duplicates}\label{sec:analysis_dup}
\vspace{-4mm}
\begin{table}[!h]
\centering
\caption{Query Results for Duplicates}
\vspace{-2mm}
\scalebox{0.78}{
\begin{tabular}{c|c|c|c|c}
\multicolumn{5}{c}{\textbf{Q1 (E = Duplicates)}} \\
\toprule
\multicolumn{2}{c|}{\textbf{R}} & \textbf{P} & \textbf{S} & \textbf{N} \\
\midrule
\multicolumn{2}{c|}{R1} & 11\% (12) & 67\% (75) & 22\% (25) \\
\multicolumn{2}{c|}{R2} & 12\% (2) & 56\% (9) & 31\% (5) \\
\multicolumn{2}{c|}{R3} & 12\% (1) & 50\% (4) & 38\% (3) \\
\bottomrule
\multicolumn{5}{c}{\textbf{Q4.1 (E = Duplicates)}} \\
\toprule
\textbf{R} & \textbf{Detection} & \textbf{P} & \textbf{S} & \textbf{N} \\
\midrule
\multirow{2}{*}{R1} & ZeroER & 5\% (3) & 61\% (34) & 34\% (19) \\
 & Key Collision & 16\% (9) & 73\% (41) & 11\% (6) \\
 \midrule
\multirow{2}{*}{R2} & ZeroER & 12\% (1) & 50\% (4) & 38\% (3) \\
 & Key Collision & 12\% (1) & 62\% (5) & 25\% (2) \\
 \bottomrule
\multicolumn{5}{c}{\textbf{Q5 (E = Duplicates)}} \\
\toprule
\textbf{R} & \textbf{Dataset} & \textbf{P} & \textbf{S} & \textbf{N} \\
\midrule
\multirow{4}{*}{R1} & Airbnb & 4\% (1) & 86\% (24) & 11\% (3) \\
 & Citation & 11\% (3) & 71\% (20) & 18\% (5) \\
 & Movie & 29\% (8) & 21\% (6) & 50\% (14) \\
 & Restaurant & 0\% (0) & 89\% (25) & 11\% (3)\\
  \bottomrule
\end{tabular}}
\label{table:result_duplicates}
\end{table}
\vspace{-2mm}




Table \ref{table:result_duplicates} shows the query results for duplicates. \textbf{(Q1):} In all the relations, there are more ``S" flags and ``N" flags than ``P" flags. Upon examining the detected duplicates, we find that duplicate detection algorithms may produce many false positives, where some non-duplicated examples are incorrectly identified as duplicates.  When these non-duplicated examples are removed from the training set, useful information may be lost, which can negatively affect the model performance.
\textbf{(Q2, Q3):}  Grouping by additional scenarios Q2 or additional ML models Q3 reveals similar findings to Q1. Hence, findings on Q2 and Q3 are omitted here.
\textbf{(Q4.1):} From results of Q4.1, we observed that in both R1 and R2, detection using ZeroER is relatively more likely to have negative impacts compared to detection with key collision method. This is because ZeroER is more aggressive and also produces more false positives than key collision detection on the datasets upon examining the cleaning results. 
\textbf{(Q4.2):}  Q4.2 is not applicable to duplicates because there is only one repair method (deletion).
\textbf{(Q5):} We only show the results of Q5 issued against R1, as the results on R2 and R3 reveal similar findings. We observed that the impact varies vastly across dataset. This is consistent with other error types, and is mainly due to the different error distributions each dataset may exhibit.


\stitle{Summary of Observations for Duplicates:} (1) Cleaning duplicates is more likely to have insignificant or negative impacts than positive impacts; (2) the impact on cleaning duplicates varies vastly across detection methods and datasets.
 

\vspace{-1mm}
\section{Overall Observations for Single Error Types}
\label{sec:conclusion}
\begin{table}[!t]
\centering
\caption{Summary of Empirical Findings for Single Error Types}
\label{T:summary_table}
\vspace{-2mm}
\scalebox{0.65}{
\begin{tabular}{c|c|c|c|c|c}
\toprule
    \multirow{2}{*}\textbf{\textbf{Error Type}} &  \multirow{2}{*}\textbf{\textbf{Impact on ML}}  & \multicolumn{4}{c}{\textbf{Does the impact depend on}}   \\ \cline{3-6}
    & & \textbf{Datasets} &\textbf{Scenarios} & \textbf{Cleaning Algos}  & \textbf{ML Algorithms}  \\
\midrule
    \textbf{Duplicates} & Varying (Mostly S \& N) & \multirow{5}{*}{Yes}  &  No & Yes & No \\
    \textbf{Inconsistencies} & Varying (Mostly S) &  & No & N.A. & No \\
    \textbf{Missing Values} & Varying (Mostly P \& S) &  &No  & Yes & No  \\
    \textbf{Mislabels} & Varying (Mostly P \& S) &  & Yes & N.A. &  No (except  Boosting) \\
    \textbf{Outliers} & Varying (Mostly S) &  &No & Yes &  No (except  KNN) \\
\bottomrule
\end{tabular}
}
\vspace{-6mm}
\end{table}

In Table~\ref{T:summary_table}, we summarize the overall findings from Sections~\ref{sec:cleanml_analysis} and discuss them in this section. 

\stitle{Strong Dependency on Dataset.} While the impact of cleaning on ML is hardly consistent, we always observe that there are noticeable differences between results of Q1 and Q5 in terms of the distributions of ``P", ``N" and ``S" impacts for all error types. This suggests that the cleaning impact depends on datasets 
 --- while two datasets may contain errors of the same type, the distributions of those errors can be vastly different.
Therefore, practitioners should never make arbitrary cleaning decisions dealing with dirty data in ML classification tasks.


\stitle{Consistent Impact from Cleaning w.r.t. Model and Scenario.}
While some ML models are noticeably more sensitive to some error types (e.g., KNN to outliers in Table~\ref{table:result_outliers} Q3 and Adaboost to mislabels in Table~\ref{table:result_mislabel} Q3) , we observe that usually there is no notable difference between results of Q1 and that of Q3 in terms of the distribution of ``P'', ``N'', and ``S'' impact, for all error types. In other words, if cleaning a dataset has a particular impact for one ML model, cleaning is likely to have the same type of impact for other models as well. This suggests that in the presence of dirty data in ML classification tasks, performing data cleaning is a more broadly applicable solution, compared with developing specific robust ML models. Similarly, the difference between the results of Q1 and that of Q2 in terms of the distribution of ``P'', ``N'', and ``S''  is also largely negligible (except for mislabels in Section~\ref{sec:mislabel_analysis}). This suggests that cleaning can be valuable in both the model development and model deployment phase.

\stitle{Strong Dependency on Cleaning Algorithms.} Because different datasets can have different error distributions (even for the same error type), no single automatic cleaning algorithm is always the best. 
%
Regardless, on all error types, we observe the same patterns comparing the results of all five queries against $R1$, $R2$, and $R3$: (1) data cleaning is more likely to have a positive impact on better ML models, i.e., models chosen by a validation set; and (2) the cleaning algorithm chosen by a validation set is more likely to have a positive impact than a randomly chosen cleaning algorithm.   In practice, selecting a data cleaning algorithm using a validation set is  sensible. However, one should not treat this as a golden solution as we do observe cases, where the selected cleaning algorithm can still degrade ML model performances (e.g.,  Table~\ref{table:result_duplicates} Q1). 

\vspace{-1mm}
\section{Mixed Errors, RobustML and Human Cleaning}
\label{sec:revision_exp}
\vspace{-1mm}
The previous section analyzes the impact of automatic cleaning methods for different single error types on various downstream ML models. In this section, we perform additional experiments to study (1) the effect of cleaning multiple error types simultaneously; (2) how does the approach of cleaning for ML compare with robust ML approaches; and (3) does human cleaning benefit downstream models more compared with automatic cleaning methods?
\vspace{-1mm}
\subsection{Cleaning Mixed Error Types}
\label{sec:mixed_errors}
\vspace{-2mm}
\stitle{Setup:} For datasets with multiple error types, we study if there is benefit in cleaning them all compared with cleaning a single error type. The cleaning algorithm space considered for cleaning multiple error types is the Cartesian product of cleaning algorithms for each component error type as listed in Table~\ref{clean_methods}. For each dataset with multiple real error types (c.f. Table~\ref{T:dataset}), we compare the best model obtained by cleaning all error types with that obtained by cleaning a single error type, i.e., the best model obtained after model selection and cleaning algorithm selection (c.f. R3 in the CleanML schema). For each dataset, we use 20 train/test splits, and perform statistical hypothesis testing, as described in  Section~\ref{sec:control_randomness}, to obtain a flag, P, S, or N --- indicating cleaning multiple error types is better  than, has no significant difference from, or is worse than cleaning a single error type, respectively.

\stitle{Results:} Table~\ref{T:mixed-error} shows the results\footnote{\scriptsize Note that we do not consider mixed error types incl.~mislabels, as we do not have datasets with coexisting real mislabels and other errors.}, where we can observe the following. (1) Cleaning all error types in a dataset is not always better than cleaning a single error type. This is similar to the impact we observe for cleaning a single error type, which does not always result in a better model compared with no cleaning and is dataset dependent. (2) The chance of having a negative impact when cleaning multiple error types is also very low. The only negative case we found is cleaning inconsistency + duplicates can be worse than cleaning inconsistency only. This is not surprising, since we discover in Section~\ref{sec:analysis_dup} that cleaning duplicates is likely to bring negative impacts. (3) On top of any single error type, additionally cleaning missing values or outliers is likely to bring positive impacts and has no negative impacts in all cases. 

\vspace{-2mm}
\begin{table}[!h]
\centering
\caption{\scriptsize Cleaning Mixed Error Types vs. Single Error Type}
\vspace{-2mm}
\label{T:mixed-error}
\resizebox{\linewidth}{!}{\begin{tabular}{c|c|c|c|c|c}
\toprule
\textbf{Dataset} & \textbf{Mixed Error Types} & \textbf{Single Error Type} & \textbf{P} & \textbf{S} & \textbf{N} \\ 
\midrule
\multirow{2}{*}{Credit} & \multirow{2}{*}{\begin{tabular}[c]{@{}c@{}}Missing Values + \\ Outliers\end{tabular}} & Outliers & 100\% (1) & 0\% (0) & 0\% (0) \\ \cline{3-6}
 &  & Missing Values & 100\% (1) & 0\% (0) & 0\% (0) \\ 
 \midrule
\multirow{2}{*}{\begin{tabular}[c]{@{}c@{}}Restaurant,\\ Movie\end{tabular}} & \multirow{2}{*}{\begin{tabular}[c]{@{}c@{}}Inconsistency + \\  Duplicates\end{tabular}} & Inconsistency & 0\% (0) & 0\% (0) & 100\% (2) \\  \cline{3-6}
 &  & Duplicates & 50\% (1) & 50\% (1) & 0\% (0) \\ \midrule
\multirow{3}{*}{Airbnb} & \multirow{3}{*}{\begin{tabular}[c]{@{}c@{}}Missing Values + \\ Outliers + \\  Duplicates\end{tabular}} & Outliers & 0\% (0) & 100\% (1) & 0\% (0) \\ \cline{3-6}
 &  & Missing Values & 0\% (0) & 100\% (1) & 0\% (0) \\ \cline{3-6}
 &  & Duplicates & 100\% (1) & 0\% (0) & 0\% (0) \\ 
 \bottomrule
\end{tabular}
\vspace{-8mm}}
\end{table}

\vspace{-5mm}
\subsection{CleanML v.s. Robust ML Approaches}
\vspace{-2mm}
\label{sec:robust_ml_vs_cleanml}
\stitle{Robust ML.} Instead of performing data cleaning to deal with noisy data, the ML community mostly focuses on developing ML algorithms robust to some particular noise type with certain distributions (e.g., \cite{quinlan1987simplifying} on noise-robust decision trees, \cite{abellan2003building} on decision trees against label noise, \cite{teng2000evaluating} on regularization to improve the robustness of ML models, \cite{khosravi2019nacl} on missing values, \cite{northcutt2017cleanlab} on mislabels). This highlights a clear advantage of data cleaning, which can be used to handle any error types and cleaned datasets can be used to train any end models without  modifications to training procedures.

\stitle{Setup:} 
As a case study, we compare CleanML with the open-source NaCL~\cite{khosravi2019nacl} package that develops a specialized form of Logistic Regression model that is robust to missing values. In addition, for all other error types, we compare CleanML with deep learning (DL) models as a form of robust ML, as they are generally less sensitive to various data errors than conventional ML models (e.g., DL robustness to massive label noise~\cite{rolnick2017deep} and to input corruptions~\cite{hendrycks2018benchmarking}). The DL model is a Multi-layer Perceptron classifier (MLP) with 
three layers and we train the model using \textit{optuna\footnote{\scriptsize We tune hidden layer size, learning rate,  momentum, and optimizer for each data split. The activation function is \textit{ReLu} and each model runs for 100 epochs. About \textit{optuna} refer to https://github.com/optuna/optuna.}}. 
For each dataset, we compare the best model obtained through data cleaning, i.e., by model and cleaning algorithm selection, with a robust end model. For each dataset, we used 20 train/test splits, and performed statistical hypothesis testing, as described in  Section~\ref{sec:control_randomness}, to obtain a flag, P, S, or N --- indicating that data cleaning is better than, has no significant difference from, or is worse than using robust ML approaches, respectively.

\stitle{Results:}  We discuss the results in Table~\ref{T:robustml-vs-cleanml}. (1) For many cases, data cleaning leads to a better end model compared with robust ML. (2) For missing values, we compare NaCL (a robust LR model) with a regular LR model under the best data cleaning algorithm, we find that cleaning is overall better (row 1). Comparing NaCL with the best model under the best cleaning algorithm, the benefit of data cleaning over robust ML widens (row 2). This exactly highlights the merit of data cleaning, i.e., data cleaning can be performed for any error type and cleaned datasets can be used to train any end model, while robust ML algorithms are usually specifically designed for an error type and a specific model type. (3) Duplicates is the only error type where deep learning is overall better than cleaning, which can be explained by our previous finding that cleaning duplicates is likely to harm ML models.






\vspace{-2mm}
\begin{table}[!th]
\centering
\caption{Robust ML vs.~Data Cleaning}
\vspace{-2mm}
\label{T:robustml-vs-cleanml}
\resizebox{0.95\linewidth}{!}{
\begin{tabular}{c|c|c|c|c|c}
\toprule
\textbf{Data Cleaning for ML} & \textbf{RobustML} & \textbf{Error Type} & \textbf{P} & \textbf{S} & \textbf{N} \\
\midrule
LR + Best Cleaning Alg & NACL & Missing Values & 33\% (2) & 50\% (3) & 17\% (1) \\
\midrule
Best Model + Best Cleaning Alg & NACL & Missing Values & 83\% (5) & 0\% (0) & 17\% (1) \\
\midrule
\multirow{4}{*}{Best Model + Best Cleaning Alg} & \multirow{4}{*}{MLP} 
   & Mislabel & 85\% (11) & 15\% (2) & 0\% (0) \\
 &  & Inconsistency & 50\% (2) & 50\% (2) & 0\% (0) \\
 &  & Outliers & 50\% (2) & 25\% (1) & 25\% (1) \\
 &  & Duplicates & 0\% (0) & 75\% (3) & 25\% (1) \\
 \bottomrule
\end{tabular}
\vspace{-2mm}}
\end{table}
\vspace{-5mm}

\subsection{Human Cleaning vs. Automatic Cleaning Algorithms}
\vspace{-1mm}
\label{sec:human_vs_automatic}
\stitle{Setup:} All cleaning algorithms considered in Table~\ref{clean_methods} are automatic cleaning algorithms in that they require no or minimal human involvement in tuning/setting up the cleaning algorithms. Here, we study if human cleaning benefits ML models compared with automatic cleaning, e.g., by spending efforts in obtaining ground-truth values for missing values or mislabels or by spending efforts in designing data quality rules for fixing inconsistencies. For each dataset, we compare the best model under human cleaning with the best model using the best automatic cleaning method. Again, we use 20 train/test splits and leverage statistical testing as described in  Section~\ref{sec:control_randomness}, to obtain a flag, P, S, or N --- indicating that human cleaning is better than, has no significant difference from, or is worse than automatic cleaning, respectively.

\stitle{Results:} (1) As we can see in Table~\ref{T:auto-vs-human}, for the two datasets ``BabyProduct" and ``Clothing", where humans manually filled missing values and corrected mislabels, the results of human cleaning are better than the best automatic cleaning method. (2) For the three datasets with inconsistencies, we manually curate data quality rules in the form of denial constraints, which are then used to detect and repair inconsistencies. We observe that the results from rule-based cleaning have no significant difference from using automatic cleaning methods.  

\vspace{-3mm}
\begin{table}[!th]
\centering
\caption{Automatic Cleaning vs.~Human Cleaning}
\vspace{-2mm}
\label{T:auto-vs-human}
\resizebox{0.95\linewidth}{!}{\begin{tabular}{c|c|c|c|c}
\toprule
\textbf{Dataset} & \textbf{Error Type} & \textbf{P} & \textbf{S} & \textbf{N} \\ 
\midrule
BabyProduct & Missing Values & 100\% (1) & 0\% (0) & 0\% (0) \\ 
Clothing & Mislabel & 100\% (1) & 0\% (0) & 0\% (0) \\ 
Company, Restaurant, University & Inconsistencies & 0\% (0) & 100\% (3) & 0\% (0) \\ 
\bottomrule
\end{tabular}}
\end{table}
\vspace{-5mm}

%% file: conclusion.tex

\section{Future Research}
\label{sec:future_research}
\vspace{-1mm}

Our study suggests that data cleaning can greatly improve downstream ML performance in many cases. This suggests that there are many opportunities for future research in the area of cleaning for ML. We share some of our thoughts.
\vspace{-1mm}
\stitle{Improving the Study.} Our current study on data cleaning for ML can be further improved/augmented in multiple ways. (1) While we focus on classification tasks, future studies could study how various errors affect other ML tasks, such as regression tasks and unsupervised clustering. (2) While we include many datasets with real-world errors for studying the impacts of automatic cleaning, more real-world datasets with ground truth could strength the study on human cleaning (e.g., one can use two versions of the same dataset, where the newer version can serve as the truth for the older one).

\vspace{-1mm}
\stitle{Theoretical Formalization.} While we provide many interesting empirical explanations for the observed impacts of cleaning on ML, it is extremely valuable to design a theoretical framework that can precisely quantify the impacts.
The DB community has long proposed the notion of \textit{consistent query answering}~\cite{arenas1999consistent}, which aims to answer SQL queries in the presence of dirty data. In consistent query answering, a tuple is included in the result of a query against the dirty database $D$ only if that tuple appears in the result of a query issued against every possible cleaned version of $D$. 
Under this framework, different classes of SQL queries and different cleaning semantics have been studied, and they entail different computation complexities for query answering. We believe that a similar theoretical framework that quantifies the impact of cleaning on ML is a fundamental step. For example, one could similarly define a notion of \textit{consistent machine learning}, where cleaning can be deemed unnecessary if the prediction of a tuple is identical for every possible cleaned version of $D$. Rigorously defining a theoretical framework for ML in the presence of dirty data is more challenging than that for SQL  --- while answering an SQL query is deterministic, the ML model training process is inherently probabilistic. 
Nonetheless, once a first attempt is taken, we expect to see many followup works that focus on different cleaning semantics and ML models.




 
\vspace{-1mm}
\stitle{Better Cleaning Solutions.} While our current study uses a fixed set of cleaning algorithms, there are ample research opportunities to design new cleaning approaches specifically for ML.
First, we could design better automatic cleaning algorithms. 
The lack of ground-truth (i.e., labeled examples) has been a long-standing challenge in designing general-purpose data cleaning solutions, which is why existing data cleaning algorithms use various proxy objectives (e.g., minimality of repairs~\cite{bohannon2007conditional}).
Fortunately, in the problem of data cleaning for ML, we have a more clearly defined objective, i.e., to improve the downstream ML model performance. Therefore, designing new cleaning algorithms that 
maximize the ML model performance is a promising direction. We anticipate the primary challenge of this direction is to avoid training the ML models many times to select the best candidate. When  multiple error types co-occur, designing automatic cleaning becomes more challenging as the space of cleaning methods becomes larger, and there may exist non-trivial interaction/dependency between different error types or cleaning methods that our current study has not considered. 
Second, we could design cleaning solutions that involve human cleaners. Our study indicates that human cleaning, especially when directly correcting data errors, can lead to even better ML models compared to automatic cleaning. However, we must also minimize/prioritize human cleaning efforts, (e.g., ActiveClean~\cite{krishnan2016activeclean} via active learning, CPClean~\cite{karlavs2020nearest} based on certain predictions), where humans are asked to clean the most beneficial examples first. 

\vspace{-1mm}